\documentclass[journal=jpccck,manuscript=article,layout=twocolumn]{achemso}
\usepackage[]{graphicx}
\usepackage{subcaption}
\usepackage[T1]{fontenc}
\usepackage[utf8]{inputenc}
\usepackage{booktabs}
\usepackage{threeparttable}
\usepackage{setspace}
\usepackage{dblfloatfix}
\usepackage{fixltx2e}
\usepackage{helvet}
\usepackage[]{amsmath}
\graphicspath{ {./Figures/} }
\author{Pedro O. Bedolla}
\email{pb@cms.tuwien.ac.at}
\affiliation[Vienna University of Technology]{Institute of Applied Physics,
Vienna University of Technology, Wiedner Hauptstra\ss e 8-10/134, 1040 Vienna,
Austria}
\alsoaffiliation[AC2T research GmbH]{Austrian Center of Competence for
Tribology (AC2T~research GmbH), Viktor-Kaplan-Stra\ss e 2, 2700 Wiener
Neustadt, Austria}
\author{Gregor Feldbauer}
\alsoaffiliation[AC2T research GmbH]{Austrian Center of Competence for
Tribology (AC2T~research GmbH), Viktor-Kaplan-Stra\ss e 2, 2700 Wiener
Neustadt, Austria}
\author{Michael Wolloch}
\alsoaffiliation[AC2T research GmbH]{Austrian Center of Competence for
Tribology (AC2T~research GmbH), Viktor-Kaplan-Stra\ss e 2, 2700 Wiener
Neustadt, Austria}
\affiliation[Vienna University of Technology]{Institute of Applied Physics,
Vienna University of Technology, Wiedner Hauptstra\ss e 8-10/134, 1040 Vienna,
Austria}
\author{Stefan J. Eder}
\affiliation[AC2T research GmbH]{Austrian Center of Competence for Tribology
(AC2T~research GmbH), Viktor-Kaplan-Stra\ss e 2, 2700 Wiener Neustadt, Austria}
\author{Nicole D\"{o}rr}
\affiliation[AC2T research GmbH]{Austrian Center of Competence for Tribology
(AC2T~research GmbH), Viktor-Kaplan-Stra\ss e 2, 2700 Wiener Neustadt, Austria}
\author{Peter Mohn}
\author{Josef Redinger}
\affiliation[Vienna University of Technology]{Institute of Applied Physics,
Vienna University of Technology, Wiedner Hauptstra\ss e 8-10/134, 1040 Vienna,
Austria}
\author{Andr\'{a}s Vernes}
\affiliation[Vienna University of Technology]{Institute of Applied Physics,
Vienna University of Technology, Wiedner Hauptstra\ss e 8-10/134, 1040 Vienna,
Austria}
\alsoaffiliation[AC2T research GmbH]{Austrian Center of Competence for
Tribology (AC2T~research GmbH), Viktor-Kaplan-Stra\ss e 2, 2700 Wiener
Neustadt, Austria}

\title{Effects of van der Waals Interactions in the Adsorption of Isooctane and
Ethanol on Fe(100) Surfaces}

\begin{document}
\begin{abstract}
  Van der Waals (vdW) forces play a fundamental role in the structure and
  behavior of diverse systems. Thanks to development of functionals that
  include non-local correlation, it is possible to study the effects of vdW
  interactions in systems of industrial and tribological interest. Here we
  simulated within the framework of density functional theory (DFT) the
  adsorption of isooctane (2,2,4-trimethylpentane) and ethanol on a Fe(100)
  surface, employing various exchange-correlation functionals to take vdW
  forces into account. In particular, this paper discusses the effect of vdW
  forces on the magnitude of adsorption energies, equilibrium geometries and
  their role in the binding mechanism.  According to our calculations, vdW
  interactions increase the adsorption energies and reduce the equilibrium
  distances.  Nevertheless, they do not influence the spatial configuration of
  the adsorbed molecules. Their effect on the electronic density is a
  non-isotropic, delocalized accumulation of charge between the molecule and
  the slab. In conclusion, vdW forces are essential for the adsorption of
  isooctane and ethanol on a bcc Fe(100) surface. 
\end{abstract}

\section{Introduction}
With a universal presence in all molecules and solids, van der Waals forces
(vdW) are important interactions to consider in the study of
matter~\cite{Lieb1986, Winterton1970, Butt2009}. Although weak in comparison to
chemical bonds, their long range nature and their collective effect play a
decisive role in the structure of molecules and their interaction with
surfaces~\cite{Atwood2004, Lawley2009, Jortner2009, Zeng2013}. Nevertheless,
their description from first principles has proven to be challenging. Most of
the common theoretical methods, such as the density functional theory (DFT),
fail to describe them properly.  Typically, the commonly used generalized
gradient approximation (GGA) underestimates the binding energies and
overestimates the equilibrium distances in various physisorbed systems, while
the ones calculated within the local density approximation (LDA) are closer to
the experiment. However, this better agreement is fortuitous since the
exponential decline of the LDA interaction cannot account for the polynomial
long-range behaviour of the vdW interactions~\cite{Mittendorfer2011,
Graziano2012}.

Over the last decade, several approximations have been proposed~\cite{Dion2004,
Kyuho2010, Sato2007, Sato2009, Sato2010, Grimme2006, Grimme2010, Grimme2011,
Jens2006, Lilienfeld2004, Lilienfeld2010, Becke2007, Tkatchenko2009,
Tkatchenko2010, Tkatchenko2012, Cole2009, Dobson1996, Furche2001, Furche2005,
Angyan2011, Hesselmann2012, Harl2009, Riley2010, Kannemann2010, Steinmann2011,
Misquitta2005, Johnson2005, Silvestrelli2008, Vydrov2010, Cooper2010} to take
dispersion into account. The most sophisticated methods aim to treat dispersion
beyond the pairwise approximation by considering collective excitations. The
many-body dispersion approach~\cite{Cole2009, Tkatchenko2012} is one of these
models that uses coupled dipoles. The dispersion interaction is obtained from
shifts in the frequencies of harmonic oscillators that occupy the atomic
positions, as the interaction between them is activated. Although promising, it
is challenging to get accurate relations between atoms and oscillator
models~\cite{Klimes2012}. Another model that includes the vdW energy accurately
is the random phase approximation~\cite{Furche2001, Angyan2011,
Hesselmann2012, Harl2009}, combined with the adiabatic connection and
fluctuation dissipation theorem~\cite{Dobson1996, Furche2005}. The
computational cost of this method limits it at present however, to small
systems commonly used as benchmarks.  Other models have been proposed that
consider dispersion to be pairwise additive but they are otherwise independent
of any external input parameters, such as VV10~\cite{Vydrov2010}, the local
response dispersion approach~\cite{Sato2009, Sato2010}, and the van der
Waals density functional (vdW-DF)~\cite{Dion2004}. Klime\v{s} and
co-workers~\cite{Klimes2010, Klimes2011} proposed a series of optimized
functionals within the framework of DFT based on the vdW-DF, which have proven
to be among the most accurate in this class of models. Furthermore, the
\mbox{vdW-DF} and its optimized versions have been implemented in extensively
distributed DFT codes and applied to a wide range of systems including
dimers~\cite{Klimes2010}, soft layered materials~\cite{Graziano2012}, organic
molecules adsorbed on graphite~\cite{Chakarova2006a, Chakarova2006b}, as well
as graphene and noble gases on metals~\cite{Mittendorfer2011, Vanin2010,
Chen2011}, among others.  These works showed the advantages and disadvantages
of these methods and consequently, increased their reliability on subsequent
applications.

Of special interest for technological purposes are the interactions between
organic molecules and metallic surfaces. The effect of non-local forces have
been previously investigated in the adsorption process occurring in promising
candidates for opto-electronic devices, including thiophene on
Cu(110)~\cite{Sony2007} and benzene, along with related compounds, on several
transition metals~\cite{Liu2012, Li2012}. The influence of dispersion forces
has also been studied in the adsorption of \mbox{\emph{n}-butane} on copper and
gold surfaces~\cite{Lee2010}, which are considered prototypes of typical weak
physisorption systems.  However, in many industrial applications the metallic
surfaces are iron-based alloys. Moreover, the involved molecules may have
larger size, be branched, or contain other types of functional groups, as in
the case of fuels and lubricants as well as their base fluids and additives.
The usefulness of these compounds strongly depends on the interacting forces at
the molecule-surface interface, since they play an important role in the
tribological behaviour, i.e., low friction and low wear are the most desired
requirements for their application. 

Recent work in the field of fuel tribology using molecular dynamics (MD)
simulations covered the stability of monolayers of stearic acid adsorbed onto
nano-rough iron surfaces during shearing~\cite{Eder2014, Eder2011} or the
frictional performance of fuel additives~\cite{Eder2013, Vernes2012}. On the
microscale, a joint numerical approach combining the finite-element method and
the boundary-element method (FEM-BEM) has been employed to simulate wear
processes in a diesel fuel lubricated sliding contact~\cite{Ilincic2013}.
However, none of these nano- and microtribological simulations using MD or
FEM-BEM explicitly considered the adsorption process as an initial step towards
the lubricity of surface-active species. Moreover, the knowledge of the
adsorption behavior of ethanol is a crucial preliminary step for a better
description of the effectiveness of additives in fuels containing
bio-components, i.e. substances with considerably higher polarity in comparison
with conventional aliphatic and aromatic fuel components. 

Last year, Tereshchuk and Da Silva~\cite{Tereshchuk2012} studied the adsorption
of ethanol and water on several transition-metal surfaces, including Fe(110).
Although the reported adsorption energies calculated with vdW corrections are
notably different from the corresponding GGA results, the vdW correction
applied in that study belongs to a class of empirical approaches which are
computationally efficient but, overall, less accurate than other available
approaches to take vdW into account~\cite{Klimes2012}. Thus, a comparison with
a method independent of external input parameters is desirable to validate the
accuracy of the calculations. Furthermore, a detailed insight into the
adsorption mechanism and the influence of the surface termination is required
for a better description of these systems. To study the effects of vdW
interactions in the adsorption process of molecules and surfaces of industrial
and tribological interest, we simulated via first principles calculations the
adsorption of isooctane (2,2,4-trimethylpentane) and ethanol on a bcc Fe(100)
surface employing various GGA and vdW functionals.  These adsorbates were
selected because isooctane is a representative aliphatic gasoline compound and
ethanol can be found---besides gasoline---in many relevant products.
Additionally, some properties differ remarkably in these molecules, such as the
chemical polarity, polarizability and chemical reactivity, and it is therefore
possible to investigate their influence on the adsorption process. This paper
describes the magnitude of adsorption energies calculated with a rigorous
treatment of non-local interactions as well as the influence of vdW forces on
equilibrium geometries and their role in the binding mechanism.

\section{Computational details}
To study the adsorption energies and equilibrium distances, we carried out
spin-polarized first-principles calculations within the framework of DFT
\cite{Kohn1965}. The Vienna Ab-initio Simulation Package (VASP)
\cite{Kresse1993, Kresse1994, Kresse1994a, Kresse1996, Kresse1996a, Kresse1999}
was used to perform the required computations. VASP produces an iterative
solution of the Kohn-Sham equations within a plane-wave basis, employing
periodic boundary conditions. The projector augmented wave (PAW) method
\cite{Blochl1994} was applied to describe the interaction between the core and
the valence electrons.

We considered various approximations to the exchange and correlation
functionals. Initially, we tested two functionals constructed in the
generalized gradient approximation (GGA): the Perdew, Burke, and Ernzerhof
(PBE) \cite{Perdew1996a, Perdew1997} and its revised version (revPBE)
\cite{Zhang1998}. These functionals differ only in one parameter of the
exchange term, \(\kappa\), changed from 0.804 in PBE to 1.245 in revPBE and
both have been extensively applied in physics and chemistry. In a second step,
we employed the vdW density functional (\mbox{vdW-DF}) by Dion et
al.,~\cite{Dion2004}
\begin{equation}
  \scalebox{0.7}{\(
  E_{\text{xc}}^{\text{vdW-DF}}\left[ n \right] =
  E_{\text{revPBE(x)}}\left[ n \right] + E_{\text{LDA(c)}}\left[ n \right]
  + E_{\text{nl(c)}}\left[ n \right],
  \)}
  \label{eq:vdw-DF}
\end{equation}
where \(E_{\text{revPBE(x)}}\left[ n \right]\) is the exchange energy obtained
with the revPBE functional, \(E_{\text{LDA(c)}}\left[ n \right]\) is an LDA
correlation and \(E_{\text{nl(c)}}\left[ n \right]\) is a non-local correlation
term which approximates the vdW interactions.  We also considered the optimized
Becke86 \cite{Becke1986} van der Waals (optB86b-vdW) functional introduced by
Klime\v{s} et al.,~\cite{Klimes2010, Klimes2011}
\begin{equation}
  \scalebox{0.7}{\(
  E_{\text{xc}}^{\text{optB86b-vdW}}\left[ n \right] =
  E_{\text{optB86b(x)}}\left[ n \right] + E_{\text{LDA(c)}}\left[ n \right]
  + E_{\text{nl(c)}}\left[ n \right],
  \label{eq:optB86b-vdW}
  \)}
\end{equation}
where a reparametrized version of the Becke86 exchange functional replaces the
revPBE exchange used in equation (1). Among the two functionals
described in equations (1) and (2), the
\mbox{optB86b-vdW} is generally more accurate \cite{Klimes2011}, and the
results obtained with it should be preferred as reference. However, we apply
the vdW-DF to analyze the effects of non-local correlation, since we can
subtract the contributions of the other terms by introducing a revPBE+LDA
functional,
\begin{equation}
  \scalebox{0.7}{\(
  E_{\text{xc}}^{\text{revPBE+LDA}}\left[ n \right] =
  E_{\text{revPBE(x)}}\left[ n \right] + E_{\text{LDA(c)}}\left[ n \right],
  \)}
  \label{revPBE+LDA}
\end{equation}
which involves terms that have been tested and applied in a wide range of
materials and whose behaviour, in contrast to the exchange term in
equation (2), is well known.

To model our systems, we constructed a supercell consisting of a body-centered
cubic (bcc) iron slab and a molecule placed on top of it. Ten layers of atoms
were included in the slab, each one containing 25 iron atoms. The vacuum
spacing in the \(z\)-direction of repeated cells was 29.73~\AA. The \(z\) axis
is parallel to the long axis of the simulation cell and starts at the bottom
layer of the iron slab. This set-up accurately models a bcc Fe(100) surface and
avoids molecule-molecule and slab-slab interactions, as well as the need to
include dipole corrections. The lattice parameter of 2.83~\AA\ used to
construct the iron slab was obtained from a bulk calculation for bcc iron,
where the calculation parameters were chosen to keep the accuracy consistent
with the rest of the computations. This value is in good agreement with the
experimental lattice constant of 2.86~\AA~\cite{Wyckoff1971}.

To ensure sufficiently accurate total energies and forces, we carefully
selected and tested our calculation parameters. The total energies were
converged to \(10^{-6}\) eV~and a cut-off energy of~400~eV was applied for
the plane-wave basis set. The k-space integrations were performed using a
\mbox{\(2\times 2\times 1\)} Monkhorst-Pack mesh \cite{Monkhorst1976,
Methfessel1989}, whereas the tetrahedron method with Bl\"{o}chl corrections
was employed for the static calculations and a Gaussian smearing with a
width of~0.2~eV for the relaxations. The conjugate gradient algorithm was
used to relax the structures, allowing the ions to move until an energy
convergence criterion of \(10^{-5}\) eV was fulfilled. 

To find the equilibrium structures, we calculated the total energies of
varied spatial configurations followed by a relaxation of the most stable one.
Various adsorption sites, slab-molecule distances and molecule orientations
were analyzed to obtain a better starting guess for the relaxations.  During
relaxations, the ions 
in the top four layers of the slab and the ones constituting
the molecule were allowed to move in all directions, while the atoms in the
remaining layers of the slab were kept fixed to simulate the bulk properties.  

We calculated the potential energy curves by varying the distance between the
molecule and iron slab, and subsequently computing the total energy of the
resulting system via static calculations. The starting geometry for the
displacements was the equilibrium structure relaxed with the corresponding
functional. The separation between the molecule and the slab (\(d\)) was
defined as the vertical distance between the atom with the lowest \(z\)
coordinate in the molecule and the iron atom closest in distance to it. The
equilibrium distance was considered to be the separation (as defined above)
between the molecule and the slab that minimizes the energy of the system. We
stress that this distance is obtained from a finite set of energy points
calculated for various values of \(d\) and therefore, its accuracy depends on
the step size between those values. 

\section{Results and discussion}

\begin{figure*}[t]
  \centering
  \begin{subfigure}[b]{0.45\linewidth}
    \center
    \includegraphics[width=0.7\linewidth]{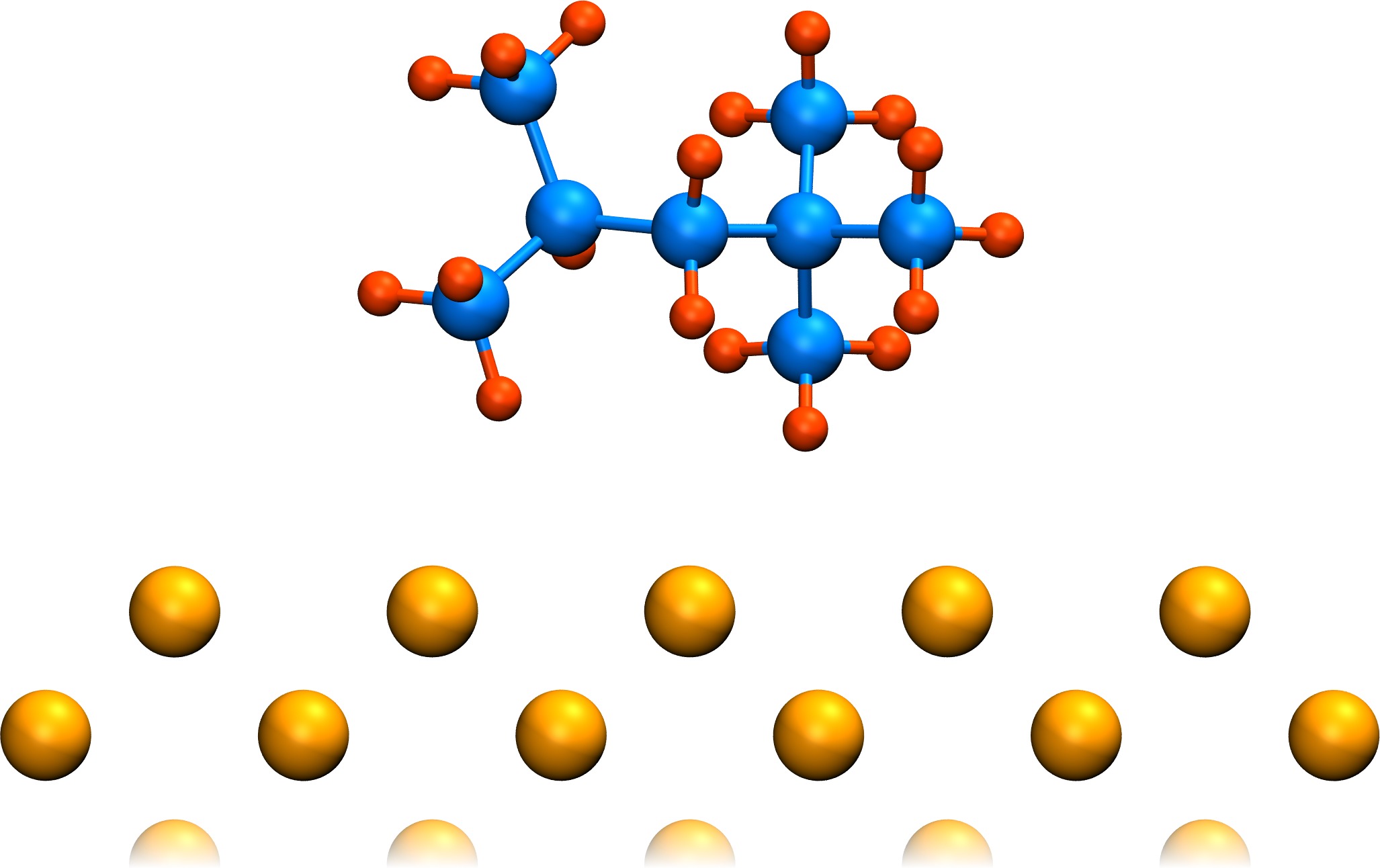}
    \caption{Side view.}
    \label{fig:fe_isooctane_vdw_side}
  \end{subfigure}
  \qquad
  \begin{subfigure}[b]{0.45\linewidth}
    \center
    \includegraphics[width=0.5\linewidth]{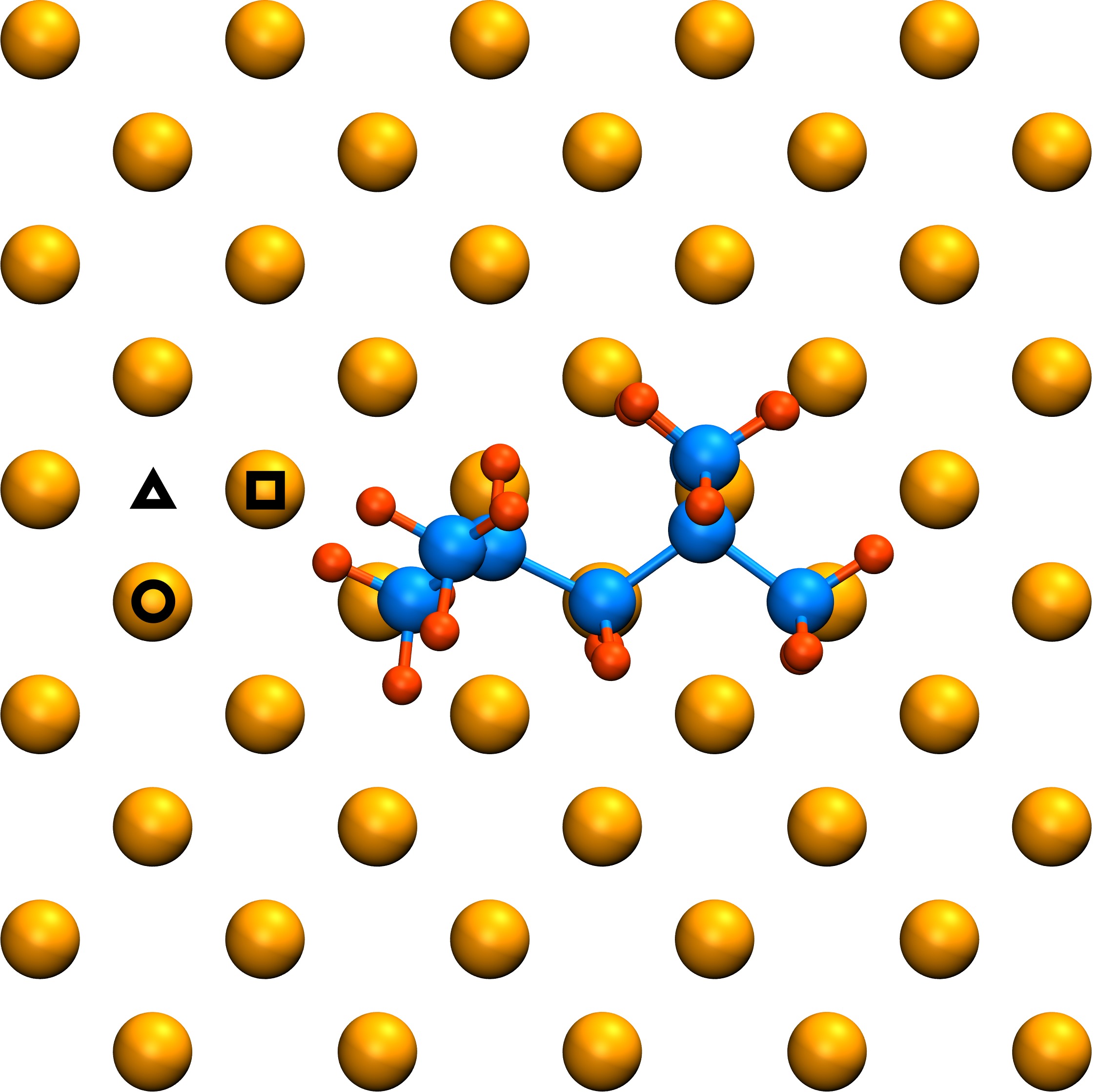}
    \caption{Top view.}
    \label{fig:fe_isooctane_vdw_top}
  \end{subfigure}
  \caption{Equilibrium adsorption geometry of an isooctane molecule on a bcc Fe(100)
  surface. The top, bridge and hollow positions are indicated by a circle,
  triangle and square, respectively.}
  \label{fig:fe_isooctane_vdw}
\end{figure*}

\begin{figure*}[b]
  \centering
  \begin{subfigure}[b]{0.45\linewidth}
    \center
    \includegraphics[width=0.7\linewidth]{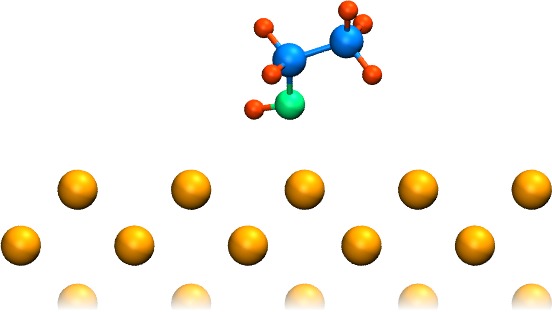}
    \caption{Side view.}
    \label{fig:fe_ethanol_vdw_side}
  \end{subfigure}
  \qquad
  \begin{subfigure}[b]{0.45\linewidth}
    \center
    \includegraphics[width=0.5\linewidth]{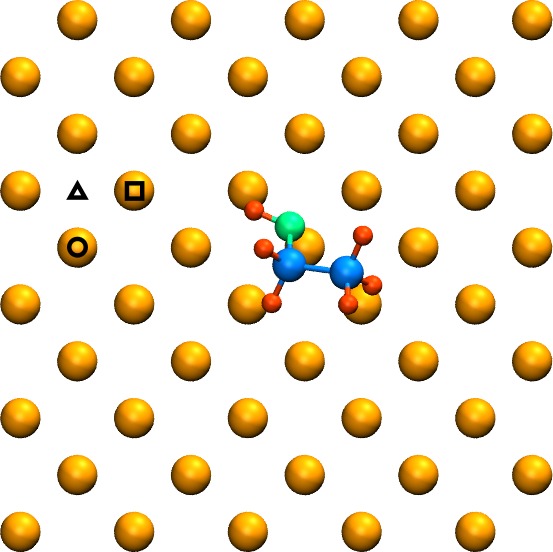}
    \caption{Top view.}
    \label{fig:fe_ethanol_vdw_top}
  \end{subfigure}
  \caption{Equilibrium adsorption geometry of an ethanol molecule on a bcc Fe(100)
  surface. The top, bridge and hollow positions are indicated by a circle,
  triangle and square, respectively.}
  \label{fig:fe_ethanol_vdw}
\end{figure*}

Neither the orientation nor the adsorption site of isooctane and ethanol is
influenced by the non-local correlation. Initially, the preferred adsorption
configurations were calculated within the GGA aproximation. After including
\mbox{non-local} interactions in our calculations via the  \mbox{optB86b-vdW}
functional, a second relaxation did not significantly change the geometry of
this configuration. The largest variations in bond lengths were of the order of
\(10^{-3}\)\AA\ and in angles, the variations were of the order of \(10^{-3}\)
degrees. In the most energetically favorable orientation, the carbon atoms of
the isooctane molecule are close to the top and hollow sites of the iron slab
(figure~\ref{fig:fe_isooctane_vdw}).  The energies of several other
orientations differ by only around 20~meV per supercell, and for this reason no
orientation is particularly favored at room temperature, where the thermal
energy \(k_{\text B}T=25\)~meV.  Similarly, in the adsorption of ethanol
several orientations of the molecule are possible. In this case, however, the
hydroxyl group always orients itself towards the slab in all low-energy
configurations (figure~\ref{fig:fe_ethanol_vdw}). The plotted geometry differs
to the one reported by Tereshchuk and Da Silva~\cite{Tereshchuk2012} for
ethanol adsorbed on the Fe(110) surface only by the C-C bond, which is almost
perpendicular to the surface. Our calculated small energy difference of 3~meV
between these two configurations indicates that both can coexist at room
temperature and that the surface termination does not influence the orientation
of the adsorbed molecule. The energy hierarchy of all PBE-structures did not
change when it was recalculated with \mbox{optB86b-vdW} to take non-local
forces into account.

\begin{figure}[h]
  \centering
  \includegraphics{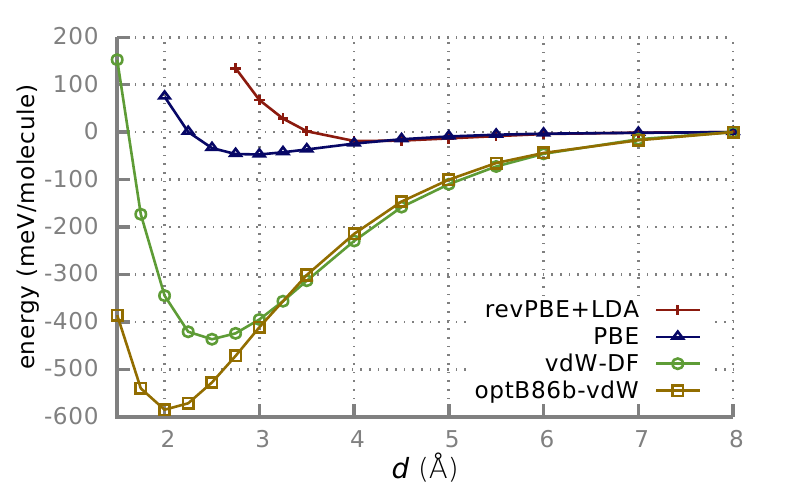}
  \caption{Calculated adsorption energy of an isooctane molecule on the bcc
    Fe(100) surface as a function of the vertical distance for the PBE,
    revPBE+LDA, vdW-DF and \mbox{optB86b-vdW} exchange-correlation potentials.}
  \label{fig:plot_fe_isooctane}
\end{figure}

Non-local interactions increase the adsorption energy of isooctane on a bcc
Fe(100) surface and reduce the equilibrium distance
(table~\ref{tab:adsorption_eqdist}). The absolute value of the adsorption
energy calculated with the \mbox{optB86b-vdW} functional is more than 12 times
larger than the one calculated with the PBE functional, while the distance
between the slab and the molecule, when in equilibrium, is reduced by 1.00~\AA\
(figure~\ref{fig:plot_fe_isooctane}). To investigate the
contribution of the non-local correlation to this increment, we compared the
adsorption energies calculated with the vdW-DF and the revPBE+LDA functionals.
The resulting energy difference is then 418~meV
(figure~\ref{fig:plot_fe_isooctane}). Since this
difference accounts for 95.8\% of the adsorption energy calculated with the
\mbox{vdW-DF} functional, the variation can now be unambiguously attributed to
the dispersion forces. By the same reasoning, the decrease in the binding distance of
1.50~\AA\ can also be associated with non-local interactions.  

\begin{figure*}[t]
  \centering
  \begin{subfigure}[b]{0.3\linewidth}
    \includegraphics[width=\linewidth]{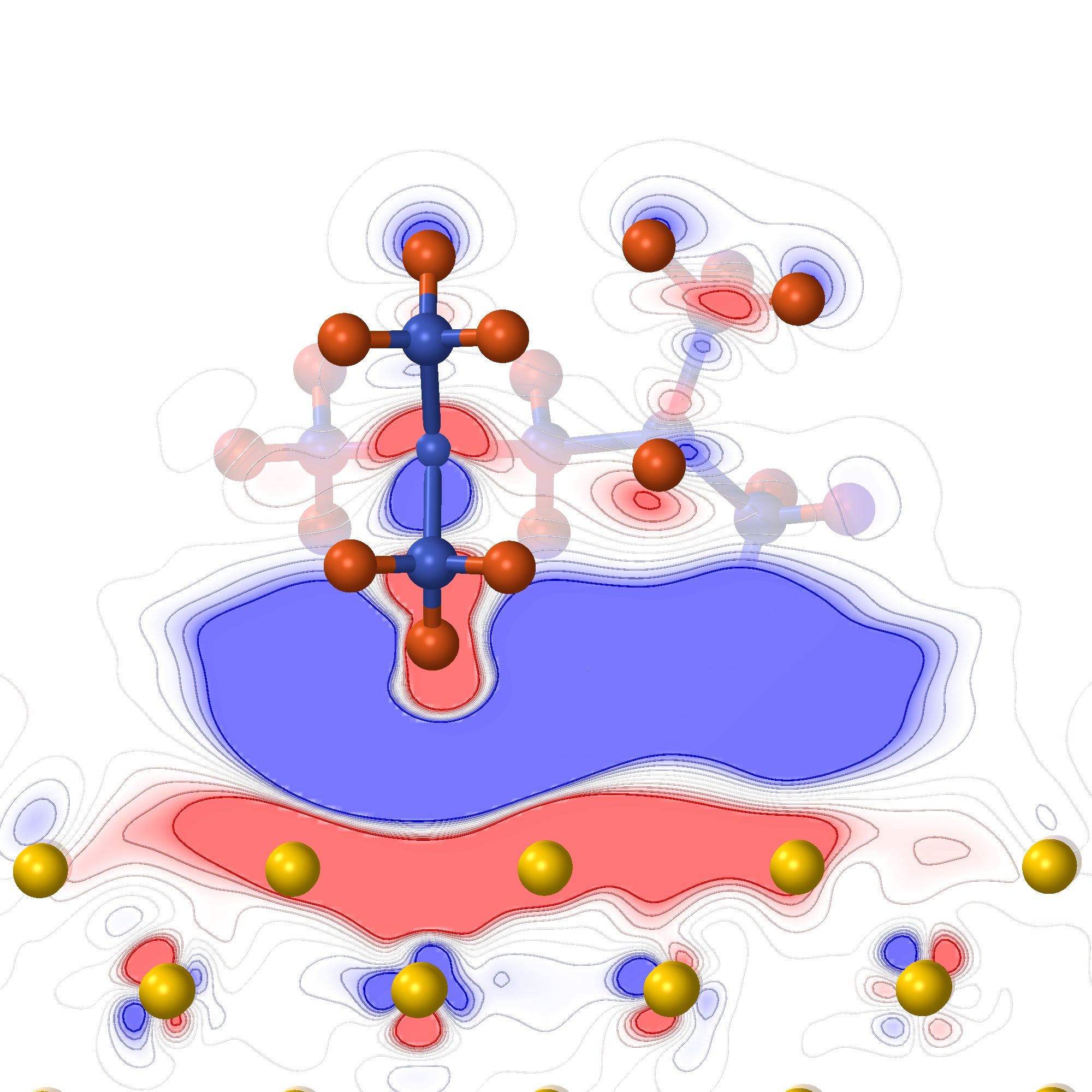}
    \caption{Calculated \(\rho_{\mathrm{diff}}\) \\ for the revPBE+LDA
    potential.}
    \label{fig:enl_isooctane_revpbe}
  \end{subfigure}
  \quad
  \begin{subfigure}[b]{0.3\linewidth}
    \includegraphics[width=\linewidth]{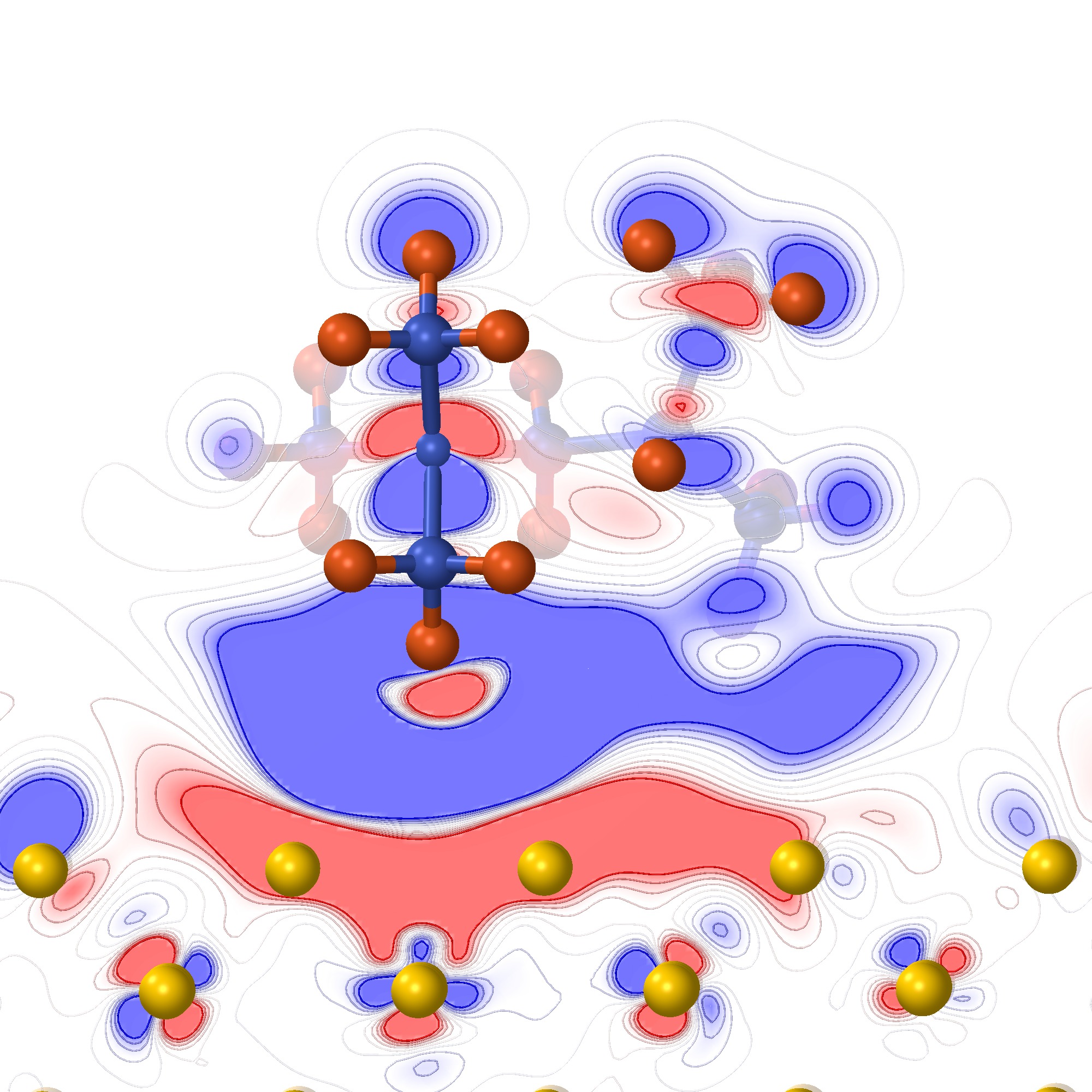}
    \caption{Calculated \(\rho_{\mathrm{diff}}\) \\ for the vdW-DF potential.}
    \label{fig:enl_isooctane_vdw}
  \end{subfigure}
  \quad
  \begin{subfigure}[b]{0.3\linewidth}
    \includegraphics[width=\linewidth]{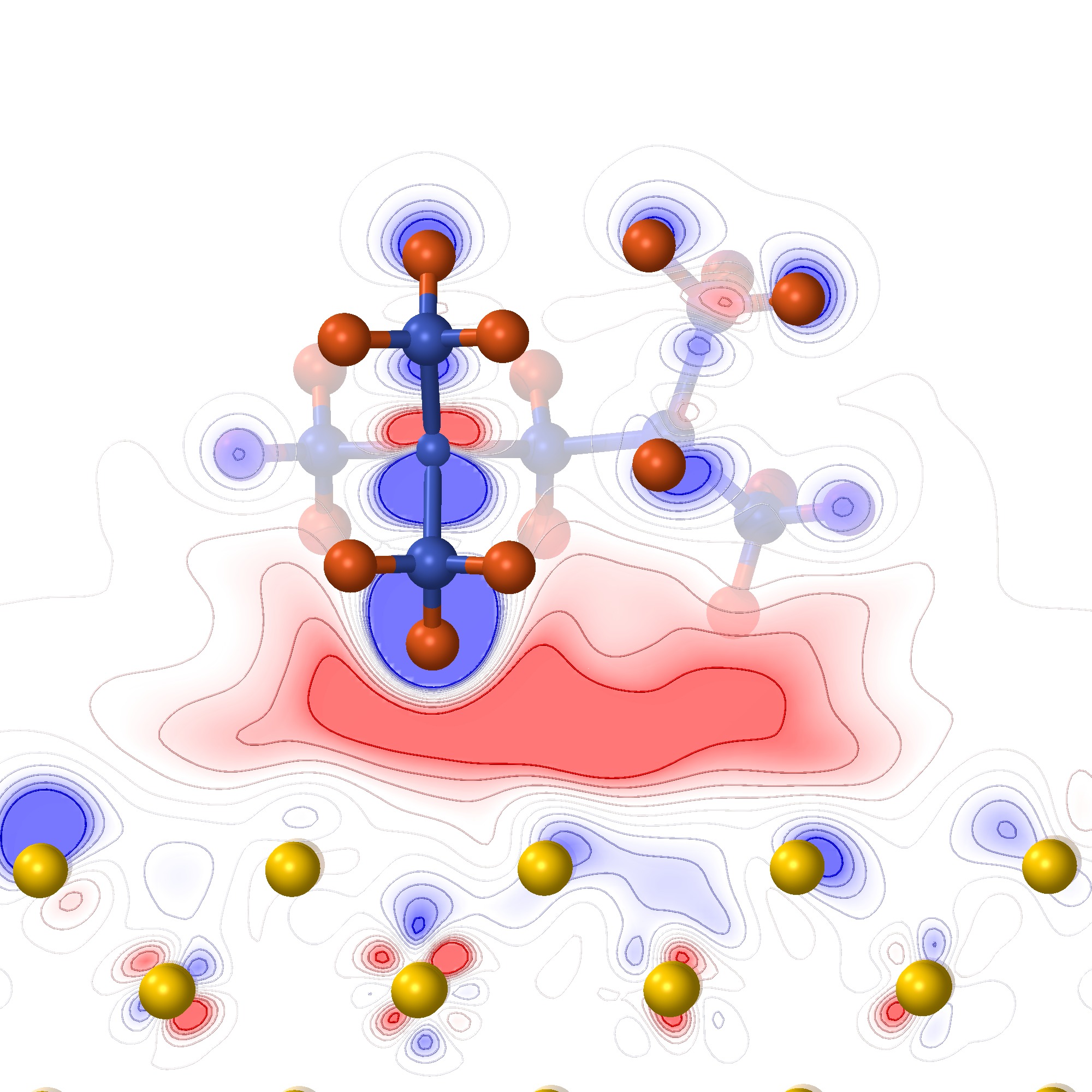}
    \caption{Subtracting (b) from (a).\\ \ }
    \label{fig:enl_isooctane_vdw_revpbe}
  \end{subfigure}
  \caption{Charge density difference (\(\rho_{\mathrm{diff}}\)) of isooctane
  adsorbed on the bcc Fe(100) surface at \(d=2.00\) \AA. The charge density
  difference is defined as
  \(\rho_{\mathrm{diff}}=\rho-(\rho_{\mathrm{isooctane}}+\rho_{\mathrm{Fe(100)}})\)
  where \(\rho\) denotes the charge density of isooctane adsorbed on Fe(100),
  while \(\rho_{\mathrm{isooctane}}\) and \(\rho_{\mathrm{Fe(100)}}\) represent
  the charge densities of the isolated molecule and the clean Fe(100) surface,
  respectively. The charge density difference is plotted in a plane
  perpendicular to the surface for values between \(-5\times10^{-4}\) (solid
  blue, deficit) and \(5\times10^{-4}\) (solid red, accumulation)
  electrons/\AA\(^{3}\).}
  \label{fig:charge_fe_isooctane}
\end{figure*}

\begin{figure*}[b]
  \centering
  \begin{subfigure}[b]{0.3\linewidth}
    \includegraphics[width=\linewidth]{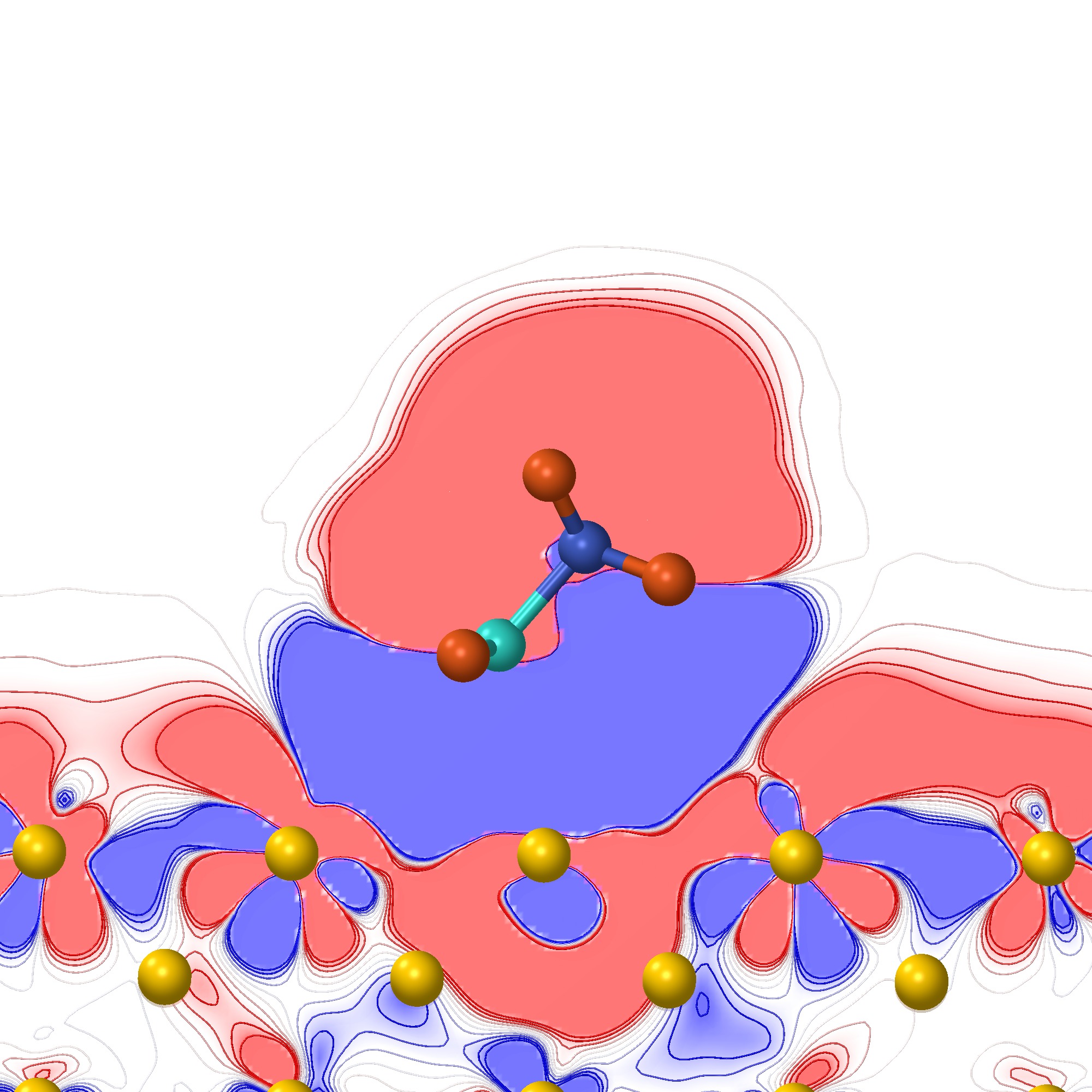}
    \caption{Calculated \(\rho_{\mathrm{diff}}\) \\ for the revPBE+LDA potential.}
    \label{fig:enl_ethanol_revpbe}
  \end{subfigure}
  \quad
  \begin{subfigure}[b]{0.3\linewidth}
    \includegraphics[width=\linewidth]{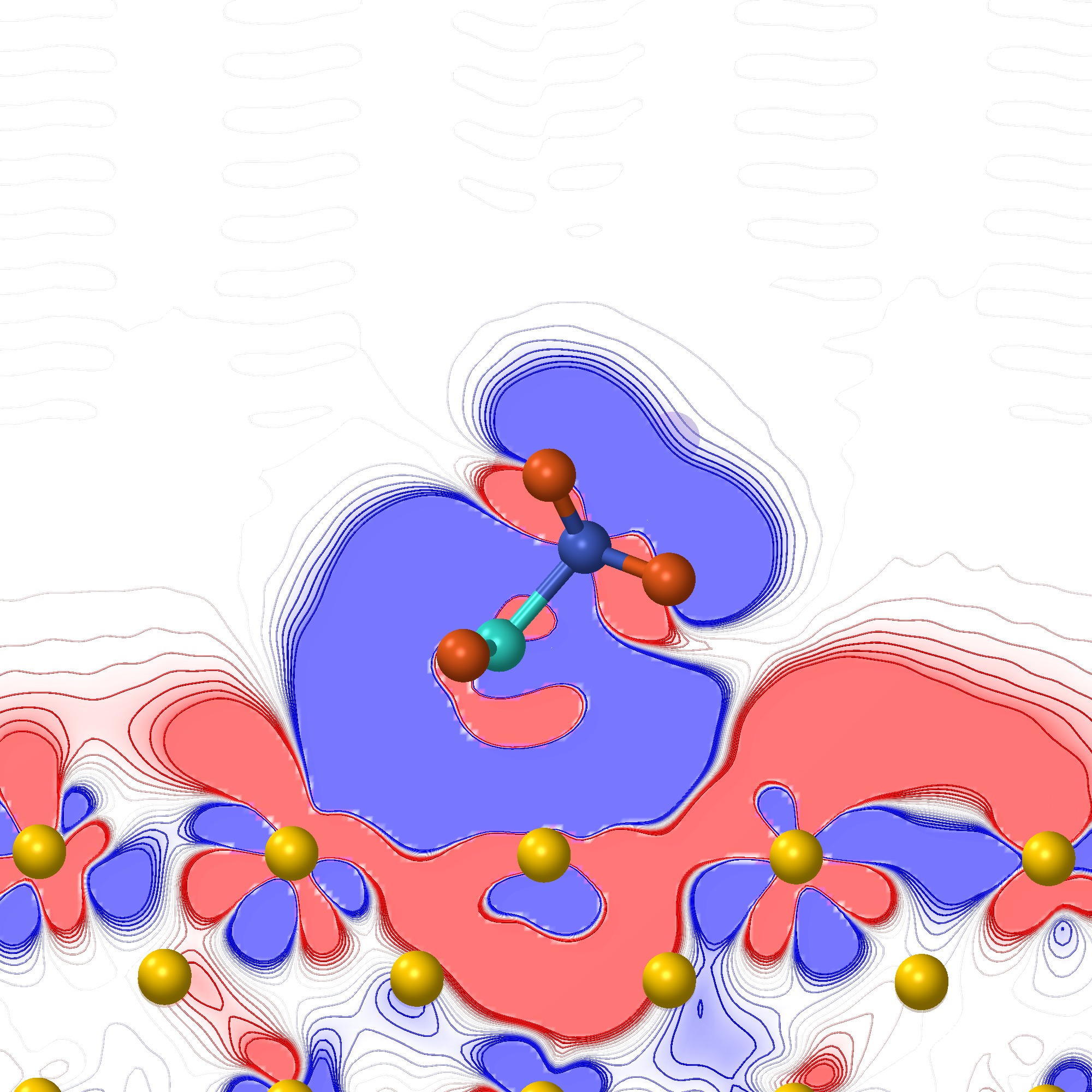}
    \caption{Calculated \(\rho_{\mathrm{diff}}\) \\ for the vdW-DF potential.}
    \label{fig:enl_ethanol_vdw}
  \end{subfigure}
  \quad
  \begin{subfigure}[b]{0.3\linewidth}
    \includegraphics[width=\linewidth]{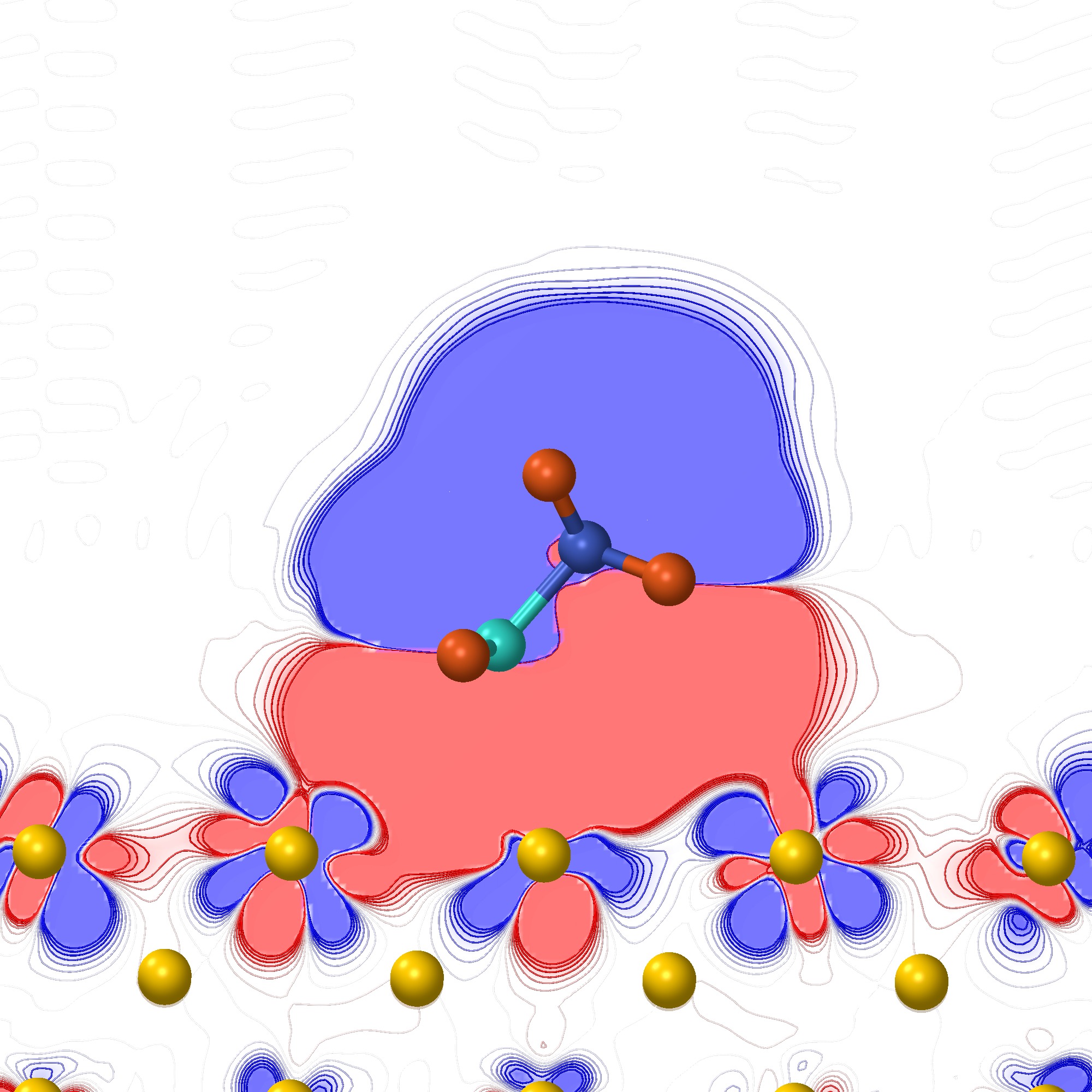}
    \caption{Subtracting (b) from (a).\\ \ }
    \label{fig:enl_ethanol_vdw_revpbe}
  \end{subfigure}
  \caption{ Charge density difference ( \(\rho_{\mathrm{diff}}\)) of ethanol
  adsorbed on the bcc Fe(100) surface at \(d=2.00\) \AA.  The charge density
  difference is defined and plotted in an analogous way to
  figure~\ref{fig:charge_fe_isooctane}.}
  \label{fig:charge_fe_ethanol}
\end{figure*}

As a result of the dispersion forces, the non-local correlation between
electrons induces the change in the adsorption energy. The equilibrium distance
is mainly determined by a balance between the long-range attractive vdW forces
and the short-range Pauli repulsion. When the isooctane molecule approaches the
iron slab, the Pauli repulsion causes a redistribution of the charge density,
particularly among the d-orbitals of the iron ions. 
No charge is transfered between the molecule and the iron slab during this
process. The overlapping between the wave functions of the molecule and the
slab accounts for the movement of the electrons to higher energy states,
increasing the total energy of the system. This effect is weaker when a proper
description of non-local interactions is considered, because the non-local
correlations reduce the electron-electron repulsion
(figure~\ref{fig:charge_fe_isooctane}). This allows the isooctane molecule to
reach a shorter equilibrium distance, where the magnitude of the attractive
forces is larger and, consequently, the binding energy increases. The
calculated equilibrium distance is also affected by the choice of the vdW
density functional, since the Pauli repulsion gives rise to exchange
interactions and these functionals differ in the description of the exchange
energy. For instance, the difference between the binding distance calculated
with the optB86b-vdW functional and the one calculated with the vdW-DF is
0.50~\AA.

\begin{figure}[h]
  \centering
  \includegraphics{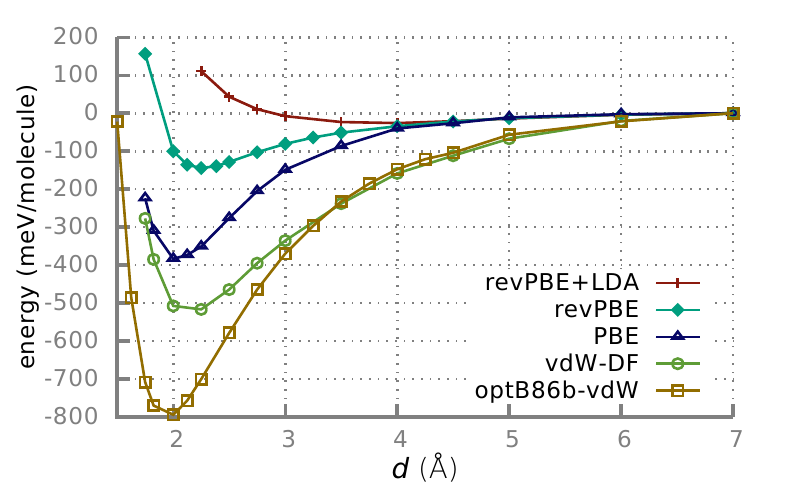}
  \caption{Calculated adsorption energy of an ethanol molecule on the bcc
    Fe(100) surface as a function of the vertical distance for the PBE,
    revPBE+LDA, vdW-DF and \mbox{optB86b-vdW} exchange-correlation potentials.}
  \label{fig:plot_fe_ethanol}
\end{figure}

As in the case of isooctane, the non-local correlation enhances the binding
energy between the adsorbed ethanol molecule and the bcc Fe(100) surface and
reduces the equilibrium distance (table~\ref{tab:adsorption_eqdist}). The
adsorption energy calculated with the \mbox{optB86b-vdW} functional is only two
times larger than the one calculated with the PBE functional, and the
equilibrium separation is 2.00~\AA\ in both cases
(figure~\ref{fig:plot_fe_ethanol}). This difference is remarkably smaller than
the one in the adsorption of isooctane and can be caused either by the
involvement of other forces in the binding mechanism or by an artifact, as in
the LDA functional. To investigate these possibilities, the effect of the
long-range interactions is extracted by comparing calculations with the vdW-DF
functional to calculations with revPBE+LDA, as previously described. The
contribution to the adsorption energy which can be attributed to the dispersion
forces is 95\% (figure~\ref{fig:plot_fe_ethanol}). Although this shows a
significant contribution of the non-local interactions to the binding
mechanism, the difference of 278~meV between the adsorption energies calculated
with the \mbox{vdW-DF} and the \mbox{optB86b-vdW} functional needs to be
investigated before discarding the contribution of other forces to
the adsorption energy.

\begin{figure*}[t]
  \centering
  \begin{subfigure}[b]{0.3\linewidth}
    \includegraphics[width=\linewidth]{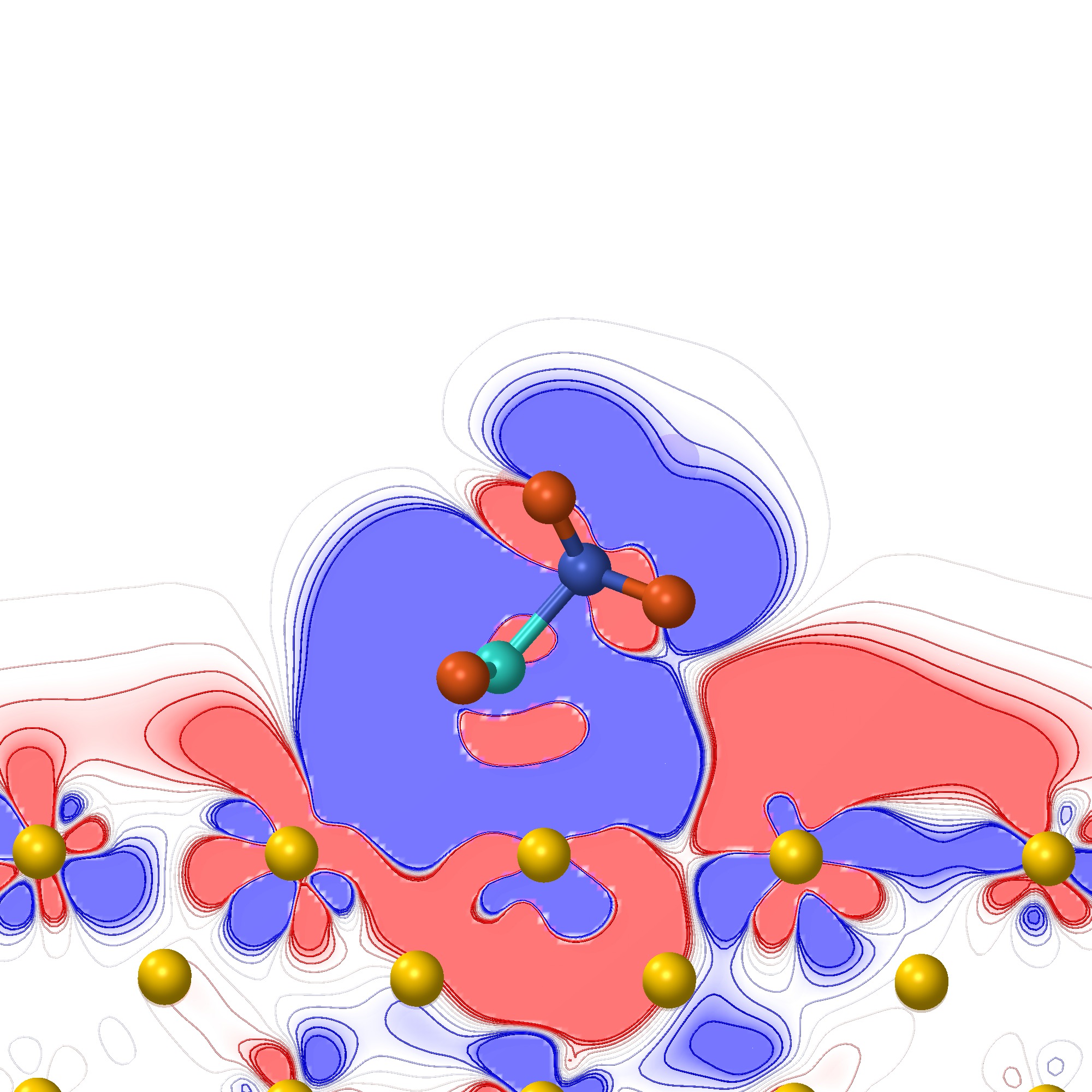}
    \caption{Calculated \(\rho_{\mathrm{diff}}\) \\ for the PBE potential.}
    \label{fig:enl_ethanol_pbe}
  \end{subfigure}
  \quad
  \begin{subfigure}[b]{0.3\linewidth}
    \includegraphics[width=\linewidth]{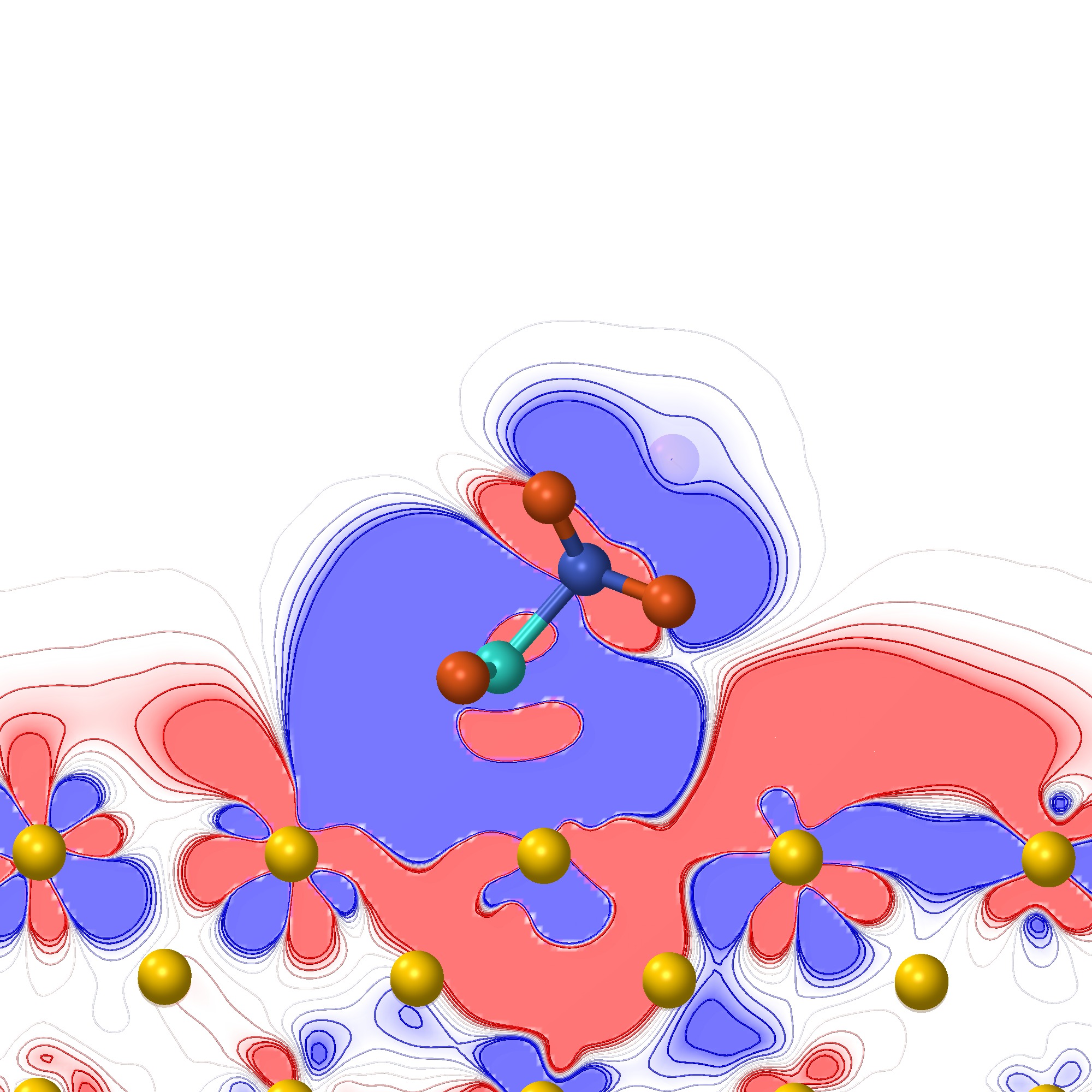}
    \caption{Calculated \(\rho_{\mathrm{diff}}\) \\ for the optB86b-vdW potential.}
    \label{fig:enl_ethanol_optb86b}
  \end{subfigure}
  \quad
  \begin{subfigure}[b]{0.3\linewidth}
    \includegraphics[width=\linewidth]{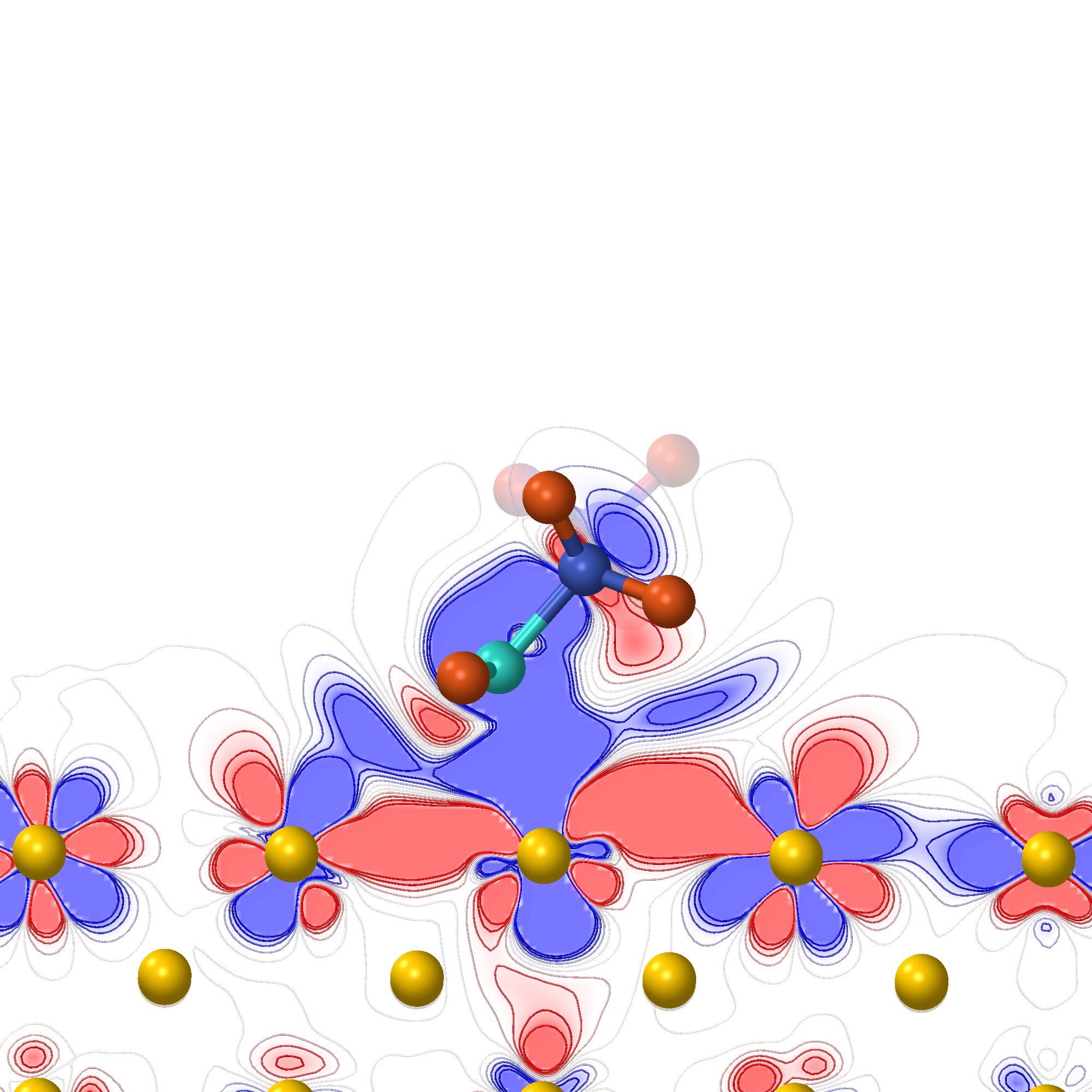}
    \caption{Subtracting (b) from (a).\\ \ }
    \label{fig:enl_ethanol_optb86b_pbe}
  \end{subfigure}
  \caption{ Charge density difference ( \(\rho_{\mathrm{diff}}\)) of ethanol
  adsorbed on the bcc Fe(100) surface at \(d=2.00\) \AA.  The charge density
  difference is defined and plotted in an analogous way to
  figure~\ref{fig:charge_fe_isooctane}.}
  \label{fig:charge_fe_ethanol_2}
\end{figure*}

\begin{figure}[h]
  \centering
  \includegraphics[width=\linewidth]{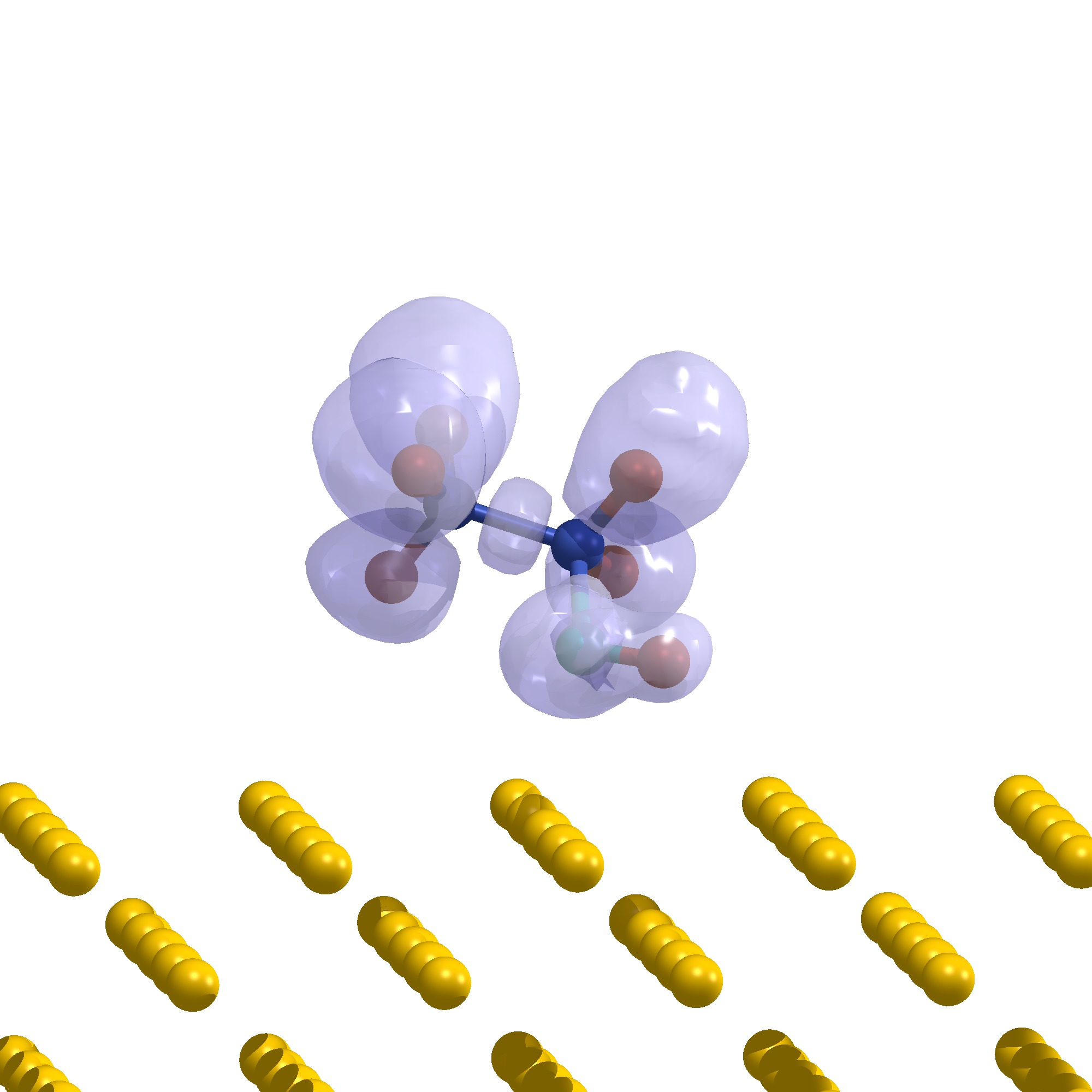}
  \caption{Isosurface at \(\text{ELF}=0.50\) of ethanol adsorbed on Fe(100) at
  \(d=2.00\) \AA\ calculated with the optB86b functional. Electrons
  ``outside'' this isosurface are expected to be delocalized.}
  \label{fig:elf_fe_ethanol}
\end{figure}

\begin{table}
  \begin{threeparttable}
    \centering
    {\scriptsize
    \begin{tabular}{lrrrr}
      \toprule
      Functional  & \multicolumn{2}{c}{Ads. energy (meV)} &
      \multicolumn{2}{c}{Eq. distance (\AA)\tnote{1}}  \\
                    \cmidrule(lr){2-3}                      \cmidrule(lr){4-5}            
                  & Isooctane & Ethanol                 & Isooctane & Ethanol \\
      \midrule
      PBE         & 47        & 384                     & 3.00      & 2.00    \\
      revPBE+LDA  & 18        & 26                      & 4.00      & 4.00    \\
      revPBE      & -         & 144                     & -         & 2.25    \\
      optB86b     & 585       & 795                     & 2.00      & 2.00    \\
      vdW-DF      & 436       & 517                     & 2.50      & 2.25    \\
      \bottomrule
    \end{tabular}
    }
    \caption{Adsorption energies and equilibrium distances calculated with
    various exchange-correlation functionals}
    \label{tab:adsorption_eqdist}
    \begin{tablenotes}
      \begin{spacing}{0.8}
      \item[1]{ \scriptsize As previously mentioned, the equilibrium distance
        is determined from a finite set of structures with fixed values of
      \(d\), with a stepsize of 0.25 \AA\ or smaller.}
      \end{spacing}
    \end{tablenotes}
  \end{threeparttable}
\end{table}

The non-local interactions contribute to the binding process of isooctane by
inducing an accumulation of charge between the iron surface and the adsorbed
molecule.  Thonhauser et al.~\cite{Thonhauser2007} showed the nature of the vdW
bonds in the argon dimer by comparing the induced electron density between two
calculations that differ in the inclusion of a term corresponding to the
non-local correlation energy. This was justified because, as a consequence of
the rapid electronic motions, the nuclei are immune to the fluctuations of the
Coulomb forces and therefore, the charge distribution must deform to produce
the required forces on the nuclei by classical Coulomb interactions alone.
These forces can be calculated, as shown by Hellman~\cite{Hellmann1937} and
Feynman~\cite{Feynman1939}, by exploiting the stationary property of the energy
with respect to variations in the wave function. Our calculations with the
vdW-DF and revPBE+LDA functional allow for a similar treatment of the molecules
considered in this work. In the adsorption of isooctane, the non-local
interactions cause a non-isotropic accumulation of charge in the region between
the molecule and the slab (figure~\ref{fig:charge_fe_isooctane}). The
electrostatic forces arising from this charge redistribution are responsible
for the adsorption of isooctane.

In addition to dispersion forces, a weak electrostatic interaction contributes
to the adsorption of ethanol on Fe(100). According to a comparison between the
results obtained with \mbox{revPBE+LDA} and \mbox{vdW-DF}, the amount of charge
that accumulates as a consequence of non-local interactions is larger and
extends over a wider area (figure~\ref{fig:charge_fe_ethanol}). However, in
contrast to the case of isooctane, in this system the charge redistributions
predicted by these functionals differ considerably. According to the results
obtained with the revPBE+LDA functional, in the absence of non-local
correlations the Pauli repulsion produces a region of deficit of charge between
the surface and the molecule and displaces the electronic density above the
molecule. This considerable redistribution of charge
(figure~\ref{fig:enl_ethanol_revpbe}) and the small adsorption energy
(table~\ref{tab:adsorption_eqdist}) calculated with this functional suggest
that it is overly repulsive in this system. A similar behavior has been
observed before for the revPBE functional (exchange and correlation) in
molecules at short separations~\cite{Murray2009}. A calculation with the revPBE
functional predicts a larger absorption energy
(figure~\ref{fig:plot_fe_ethanol}) but it is, nevertheless, less than half of
the one calculated with the PBE functional. Evidently, this system is heavily
influenced by the description of the exchange energy, since the only difference
between these two functional lies in the exchange term. The charge
redistribution calculated with the PBE functional
(figure~\ref{fig:enl_ethanol_pbe}) resembles the one calculated with vdW
functionals (figures~\ref{fig:enl_ethanol_vdw} and
\ref{fig:enl_ethanol_optb86b}). Since PBE does not include the effects of
non-local correlations, this last result indicates that an important
contribution to the binding between the ethanol and the Fe(100) surface cannot
be attributed to the dispersion forces alone. Based on previously estimated
adsorption energies, it has been proposed that a weak chemisorption is involved
in the binding mechanism~\cite{Tereshchuk2012}. Our calculations, however, do
not show a charge transfer large enough to consider the formation of an ionic
bond and the analysis of the electron localization function
(ELF)~\cite{Becke1990, Silvi1994} do not point to the existence of localized
electrons forming a bond between the molecule and the metallic slab
(figure~\ref{fig:elf_fe_ethanol}). The optB86b-vdW functional predicts a lesser
amount of charge between the ethanol and the iron slab than the one calculated
with PBE, particularly between the hydroxyl group and the closest iron atom to
it (figure~\ref{fig:enl_ethanol_optb86b_pbe}). Since this effect is accompanied
by an increase in the adsorption energy, a weak electrostatic interaction
together with the dispersion forces may constitute the main mechanisms
contributing to the adsorption. Nevertheless, this comparison does not allow to
separate these contributions, since these functionals differ in the exchange
term.

The difference in the adsorption process of isooctane and ethanol on a bcc
Fe(100) surface can be understood in terms of the polarizability of the
molecules and the charge density distribution around the functional group.
Clearly, isooctane is expected to be more polarizable than ethanol because of
its larger molecular size.  Additionally, it is well known that alkanes are
among the most polarizable molecules \cite{Anslyn2006}. Since dispersion forces
arise from the formation of instant multipoles, their strength is directly
related to the polarizability. Moreover, the absence of a functional group in
isooctane leaves dispersion forces as the only possible binding mechanism.  On
the other hand, ethanol has a permanent dipole, product of the difference in
electronegativity between the hydrogen and the oxygen atom in the hydroxyl
group, and non-bonding electrons in the oxygen. During the adsorption process,
not only the dispersion forces contribute to the binding but there is also a
weak electrostatic interaction between the hydroxyl group and the iron slab,
where the charge redistribution leads to the formation of multipoles on the top
layer (figures \ref{fig:charge_fe_ethanol} and \ref{fig:charge_fe_ethanol_2}).
Consequently, the iron surface binds ethanol stronger than isooctane even
though the contribution of the dispersion forces is expected to be weaker than
in isooctane.

\section{Summary and outlook}
The vdW forces are essential for the adsorption of isooctane and ethanol on a bcc
Fe(100) surface. As product of these long-range interactions, the non-local
correlation leads to an increase in the adsorption energies and a reduction of
the equilibrium distances. Nevertheless, they do not influence the spatial
configuration of the adsorbed molecules. Their effect on the electronic
density is a non-isotropic accumulation of charge between the molecule and the
slab.

Our calculations are a first approach towards a more rigorous treatment of vdW
interactions in complex systems and they will contribute, after experimental
validation, to the development and improvement of vdW functionals which are
independent of external input parameters. We showed the effects of non-local
interactions in the electronic density and in the adsorption mechanism.
Isooctane binds to the Fe(100) surface via dispersion forces while in ethanol,
in addition to the dispersion forces, a weak electrostatic interaction between
the hydroxyl group and the iron surface contributes to the binding.

We anticipate that with the continuous increase in computing power, future
calculations that consider many-body effects in combination with the work here
presented will clarify the relevance of these interactions and contribute to a
more accurate analysis of the vdW forces in systems of industrial interest. In
particular, the obtained results will aid the fitting of accurate
surface--lubricant interaction potentials required for classical MD simulations
including numerous organic molecules.  These interface potentials usually
constitute a considerable uncertainty in any such MD simulation, as they are
rarely assumed more advanced than a Lennard-Jones potential parameterized
according to desorption data. A detailed knowledge of the interaction energy
between molecules and metal surface can therefore greatly boost the precision
of large-scale atomistic studies of the thermal, mechanical and structural
stability of molecular surface films.

\begin{acknowledgement}
  POB, JR and PM acknowledge support from the Austrian Science Funds (FWF)
  within the SFB ViCoM F4109-N13 P09.

  Part of this work was funded by the Austrian COMET Program (Project K2
  XTribology, Grant No. 824187), the ERDF, and the province of Lower Austria
  (Onlab Project) and was carried out at the ``Excellence Centre of
  Tribology''.

  The computational results presented have been achieved in part using the
  Vienna Scientific Cluster (VSC).
\end{acknowledgement}


\end{document}